# Tweeting about journal articles: Engagement, marketing or just gibberish?[1]

Nicolas Robinson-Garcia[*], Rakshit Trivedi[**], Rodrigo Costas[***], Kimberley Isset[****], Julia Melkers[****], and Diana Hicks[****]

[*] elrobin@ingenio.upv.es
INGENIO (CSIC-UPV), Universitat Politècnica de València, Valencia (Spain)

[**] rtrivedi6@gatech.edu
College of Computing, Georgia Institute of Technology, Atlanta (United States)

[***] rcostas@cwts.leidenuniv.nl
Centre for Studies of Science and Technology (CWTS), Leiden University, Leiden (The Netherlands)

[****] kim.isett@pubpolicy.gatech.edu; julia.melkers@pubpolicy.gatech.edu; dhicks@gatech.edu
School of Public Policy, Georgia Institute of Technology, Atlanta (United States)

## INTRODUCTION

Altmetric indicators, defined as mentions from social media platforms to scientific literature, are seen as promising and with the potential to be used, when fully developed in research evaluation (Wilsdon et al., 2015). They are perceived as a means to connect with 'broader audiences' translating research findings from academia to society. Despite this enthusiastic appraisal on what could become from using social media in research evaluation, current findings leave mixed feelings due to the incapability to comprehend fully what we are analysing or measuring – societal impact, social engagement, interest? Research studies in this area have mainly focused on establishing comparisons with citation indicators (Haustein et al., 2014; Costas, Zahedi & Wouters, 2015), their disciplinary coverage (Zahedi, Costas & Wouters, 2014), or differences between the social media platforms altmetric data providers offer (Robinson-Garcia et al., 2014).

Among other findings, we highlight the dominance of Twitter data in altmetrics (Robinson-Garcia et al., 2014), the volatility of the metrics (Torres-Salinas, Cabezas-Clavijo & Jiménez-Contreras, 2013) or the heterogeneity of sources (Haustein, 2016). These findings have led to conclude that in many cases, social media mentions to literature are just another informal communication channel for researchers rather than a means to engage with non-academics (Sugimoto et al., 2016), questioning if the approach of translating the citation analogy to social media is the most appropriate way to capture social interactions between scholars and lay people (Robinson-Garcia, van Leeuwen & Rafols, 2017).

---

[1] Nicolas Robinson-Garcia has received support from a Juan de la Cierva-Formación postdoctoral fellowship from the Spanish Ministry of Economy and Competitiveness. He also received funding the Fulbright Commission and José Castillejo. This work was partially supported by NIH grant U19-DE-22516 and NSF award number 1445121.

In this work in progress we focus on the use of one particular social media platform: Twitter. This platform has been largely analysed due to its larger coverage of mentions to scientific papers as well as to the keen interest shown by many researchers on the use of Twitter to engage with other audiences. Here we find a genuine interest in the field of biomedicine, one of the few to first suggest the use of Twitter mentions or 'tweetations' to predict citations (Eysenbach, 2011). Also, Twitter has been described as a transformative tool in health care (Hawn, 2009), connecting with colleagues and patients (Alpert & Womble, 2016), complementing traditional teaching methods (Nason et al., 2015) or critically appraising and reviewing research (Maclean et al., 2013). This latter motivation is the one behind the 'altmetrics manifesto' (Priem et al., 2010), when claiming that "altmetrics will track impact outside the academy, impact of influential but uncited work". It is also 'appraisal' the function that Haustein et al. (2016a) suggest Twitter mentions to journal articles play. Still, issues such as the existence of bots (Haustein et al., 2016b) or the automatic nature of the content of tweets suggest otherwise.

The study presented here is part of a larger study on the transmission of scientific literature to professional practice in the field of Dentistry. As such, our goal is to understand the role played by Twitter (if any) in such enterprise. To do so, we present two case studies in which we critically examined the contents and context of tweets linking to scientific papers in the field of Dentistry. The first case study focuses on the analysis of tweet contents from users from the United States. We adopt a qualitative approach and examine the tweets of the top 10 most tweeted papers and consider the motivations of the users behind them as well as the nature of such tweets. In the second case study, we focus on the role journals play on the total number of Twitter counts their publications receive and analyse their capability to influence and engage audiences on their contents.

## METHODOLOGICAL APPROACH
We retrieved all publications from all years of 84 journals in the field of Dentistry indexed in Web of Science in 2016. Additionally, we included the publications of 47 other journals indexed in PubMed also in the field of Dentistry. We obtained a total of 196,812 papers which we then crossed with the Altmetric.com API through their PMID identifier in June 2016. 15,894 publications were found to have been mentioned in a total of 52,540 tweets between 2011 and 2016. 2,202 tweets reported to originate from US accounts. This subset is the one analysed in our first case study.

We then manually identified for the complete dataset all accounts associated to scientific journals. We performed a descriptive analysis on the number of tweets the produce and the number of retweets they receive and compared this with the total number of tweets mentioning their research articles. This analysis is the one described in our second case study.

## CASE 1. THE DENTAL CONVERSATION IN THE US
In this case study, we will show how most of the tweets of the top tweeted papers are either run by monomania and single-issue campaigner, social media managers and the influence of journals themselves on promoting their own contents. Table 1 shows a summary of our findings when focusing on the top 10 most tweeted dental journals. We will also discuss the lack of 'humanity' in the contents of tweets as well as the influence of bots on the counting of tweets, questioning the capacity of Twitter as a social media platform to engage on scientific discussions with academic and non-academic audiences.

**Table 1. Top 10 most tweeted dental papers in the US**

| Paper title | Explanation | Publication year | WoS cites | Tweets | Accounts | Tweet variants | Begins with @ |
|---|---|---|---|---|---|---|---|
| Acetaminophen Old Drug, New Issues. | Single-issue campaigner | 2015 | 9 | 264 | 15 | 71 | 103 |
| Dietary Carbohydrates and Dental-Systemic Diseases | Single-issue campaigner | 2009 | 36 | 70 | 17 | 30 | 14 |
| Fluoridation and social equity. | #oralhealthequity | 2002 | 42 | 59 | 41 | 4 | 0 |
| From victim blaming to upstream action: tackling the social determinants of oral health inequalities | #oralhealthequity | 2007 | 159 | 54 | 33 | 3 | 0 |
| The Effect of Waxed and Unwaxed Dental Floss on Gingival Health: Part I. Plaque Removal and Gingival Response | Social media manager | 1982 | 17 | 51 | 44 | 2 | 0 |
| Electronic cigarettes induce DNA strand breaks and cell death independently of nicotine in cell lines | Largely single-issue campaigner | 2016 | 12 | 39 | 34 | 13 | 2 |
| The tooth-worm: historical aspects of a popular medical belief. | Social media manager | 1999 | NA | 39 | 39 | 3 | 0 |
| Oral Health Literacy among Female Caregivers: Impact on Oral Health Outcomes in Early Childhood | | 2010 | 47 | 35 | 25 | 7 | 0 |
| Why do GDPs fail to recognise oral cancer? The argument for an oral cancer checklist | Retweets of BDJ tweets | 2013 | 6 | 29 | 18 | 13 | 0 |
| Beyond the DMFT: the human and economic cost of early childhood caries. | | 2009 | 103 | 28 | 25 | 2 | 0 |

We argue conceptual problems when trying to establish an analogy on the meaning of tweeting research articles in comparison with citing in a journal article another paper.

## CASE 2. THE ROLE OF JOURNALS ON PROMOTING THEIR PUBLICATIONS

In this second case study, we focus on the effect of journals' activity in Twitter on the overall number of tweets papers receive. For this we identified 20 Twitter accounts associated to journals. These accounts belonged either to the journals themselves, the scientific associations or societies behind them or their publishers. Their tweets represented 23.7% of the total of 52,540 tweets retrieved. Table 2 shows the number of tweets each type of journal-related account produced, share of tweets linking to their own journals (self-tweets), number of retweets from other accounts, total number of impressions (that is, number of retweets from other accounts of their tweets) and uptake rate (impressions/tweets).

**Table 2. Number of tweets by type of journal-related account**

|  | journal | scientific association | publisher |
|---|---|---|---|
| # tweets | 11825 | 561 | 85 |
| % self-tweets | 97,99% | 83,78% | 98,82% |
| # retweets | 59 | 218 | 16 |
| # impressions | 9230 | 695 | 60 |
| Uptake rate | 0,78 | 1,24 | 0,71 |

Figure 2 offers a first glimpse on the differences between journals on their strategy and success at promoting their contents. Among others, the British Dental Journal stands out as a journal highly devoted to promoting their contents through Twitter.

**Figure 2 Independent and journal related tweets and journal citations**

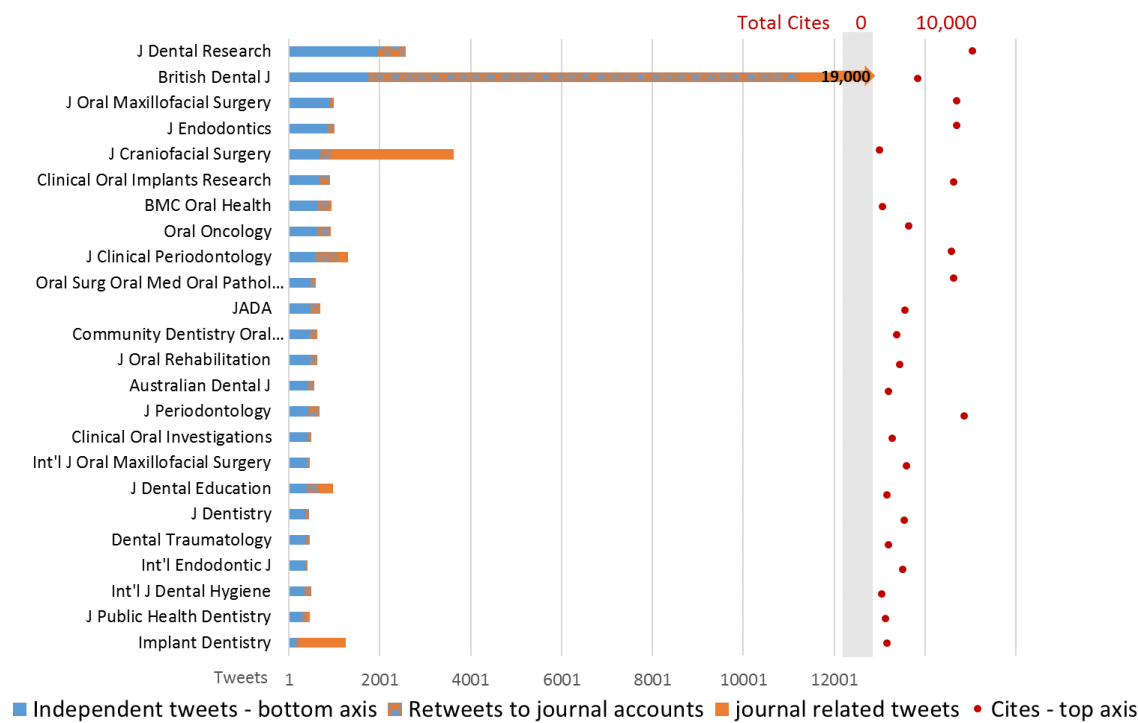

## CONCLUDING REMARKS

This paper presents preliminary results on the analysis of tweets to journal articles in the field of Dentistry. We present two case studies in which we critically examine the contents and context that motivate the tweeting of journal articles. We then focus on a specific aspect, the role played by journals on self-promoting their contents and the effect this has on the total number of tweets their papers produce. In a context where many are pushing to the use of altmetrics as an alternative or complement to traditional bibliometric indicators. We find a lack of evidence (and interest) on critically examining the many claims that are being made as to their capability to trace evidences of 'broader forms of impact'. Our first results are not promising and question current approaches being made in the field of altmetrics.